\documentclass{ws-procs975x65}

\begin{document}

\title{QUANTIZATION OF THE MASSLESS SCALAR FIELD \\ IN  DE SITTER SPACETIME WITH UNITARY DYNAMICS}

\author{J. CORTEZ}
\address{Departamento de F\'\i sica,
Facultad de Ciencias, Universidad Nacional Aut\'onoma de M\'exico,\\
M\'exico D.F. 04510, Mexico\\
E-mail: jacq@ciencias.unam.mx}

\author{D. MART\'IN-DE BLAS$^*$ and G. A. MENA MARUG\'AN$^{**}$}

\address{Instituto de Estructura de la Materia,
CSIC, Serrano 121, 28006 Madrid, Spain
\\
$^{*}$E-mail: daniel.martin@iem.cfmac.csic.es, $^{**}$E-mail: mena@iem.cfmac.csic.es}

\author{J. M. VELHINHO}

\address{Departamento de F\'\i sica, Universidade da Beira
Interior, 6201-001 Covilh\~a, Portugal\\
E-mail: jvelhi@ubi.pt}


\bodymatter
\section{Introduction}

It  was recently proved \cite{cmv,cmov} that a quantization with unitary dynamics exists and is essentially unique, for free scalar fields in every  spacetime that is conformal to a static spacetime with compact spatial sections of dimension no greater than 3, in the case where the conformal factor is exclusively time dependent.

This applies to the propagation of fields in spatially compact FRW models (and to the quantization of high frequency modes in open models), but also to a great variety of physically unrelated cases, including the quantization of purely gravitational degrees of freedom, such as those of the inhomogeneous cosmologies know as Gowdy models,
whose effective dynamics is of the above type\cite{ccmv}.

This general result has been questioned by a claim that it fails to apply to the massless scalar field in de Sitter spacetime\cite{vv}. The massless case is indeed special in that no well defined de Sitter invariant vacuum exists, i.e.~there is no Bunch-Davies vacuum in the massless case. However, this is no obstruction to the existence of a well defined quantization with unitary dynamics in this case, in contradiction with the conclusions of Ref.~\refcite{vv}. Following Ref.~\refcite{cbmv}, we present here a Fock quantization with unitary dynamics for the massless field in de Sitter spacetime. It turns out that the quantization obtained from our approach is unitarily equivalent to the O(4)-invariant quantization introduced by Allen and Folacci\cite{af}.

\section{Quantization of the massless scalar field with unitary dynamics}

The de Sitter spacetime is the maximally symmetric spacetime of positive constant curvature. Using conformal time, the metric of spacetime can be written in the form 
\begin{equation}
ds^2= {{a^2(\eta)}}[-d\eta^2+d{\sigma}^2+\sin^2({\sigma})(d\theta^2+\sin^2(\theta)d\varphi^2)],
\end{equation}
where one recognizes the metric of the static universe $\mathbb{R}\times S^3$ and the exclusively time-dependent conformal factor:
\begin{equation}
\label{scale}
a^2(\eta)=12R^{-1}\sin^{-2}(\eta),
\end{equation}
where $R$ is  the curvature of spacetime.

Let us consider the propagation of a free (minimally coupled) massless real scalar field $\phi$.  Using conformal time and introducing the scaled field
\begin{equation}
\label{rescale}
{{\chi=a(\eta) \phi}},
\end{equation}
one arrives at the field equation
\begin{equation}
\label{feq}
\ddot\chi-[\Delta {-1}+R{a^2}/6]\chi=0,
\end{equation}
where $\Delta$ is the Laplace-Beltrami operator on $S^3$ and the dot stands for the derivative with respect to the conformal time $\eta$. Up to a total time derivative, the Lagrangian density is
\begin{equation}
L=\frac{1}{2}\left[\left(\dot{\chi}\right)^2 -\left(\nabla\chi\right)^2 +\frac{\ddot{a}}{a}\,\chi^2\right],
\end{equation}
to which corresponds the canonical momentum conjugate to $\chi$:
\begin{equation}
\label{momentum}
P_\chi= \dot{\chi}.
\end{equation}
Let us now decompose the field  $\chi$ and the momentum $P_\chi$ in terms of the usual spherical harmonics, to arrive at the Hamiltonian equations of motion for harmonic modes:
\begin{equation}
\dot q_{k\ell m} = p_{k\ell m}, \qquad \dot p_{k\ell m} = \left[2\sin^{-2}\eta-(k+1)^2 \right]q_{k\ell m}.
\end{equation}
The general solution to the Hamiltonian equations for the modes can be written in terms of the associated Legendre functions $P^\mu_\nu$ and $Q^\mu_\nu$, whose properties (including asymptotic properties) are well known (see Ref.~\refcite{cbmv} for details). In particular, one can check that
\begin{equation}
\label{soluc}
q_{k\ell m}(\eta)=A_{k\ell m}\sqrt{\sin\eta}P^\mu_\nu(-\cos\eta)+
B_{k\ell m}\sqrt{\sin\eta}Q^\mu_\nu(-\cos\eta),
\end{equation}
where $A_{k\ell m}$ and $B_{k\ell m}$ are  constants, $\nu=k+1/2$ and  {$\mu=3/2$}.

To show the existence of a Fock quantization with unitary dynamics, let us introduce the classical (complex) variables
\begin{equation}
\label{cavar}
\displaystyle
a_{k\ell m}={\frac{1}{\sqrt{2\omega_k}}} \left(\omega_k q_ {k\ell m}+i p_{k\ell m}\right), \qquad a^*_{k\ell m}={\frac{1}{\sqrt{2\omega_k}}} \left(\omega_k q_{k\ell m}-i p_{k\ell m}\right),
\end{equation}
where  $\omega_k{=}k+1$.

If it is now declared that the  variables $a_{k\ell m}$ and $a^*_{k\ell m}$ are to be quantized as creation and annihilation operators, one is defining a particular Fock quantization for the system. [In other words, the complex structure $J$ that determines the Fock quantization is given by $J(a_{k\ell m})=ia_{k\ell m}$, $J(a^*_{k\ell m})=-ia^*_{k\ell m}$.]

In terms of the variables (\ref{cavar}), classical evolution between an initial time $\eta_0$ and final time $\eta$ is of the standard Bogoliubov form
\begin{equation}
\label{evolution-a}
\left(\begin{array}{c} a_{k\ell m}(\eta) \\ a^*_{k\ell m}(\eta) \end{array}\right)=\left( \begin{array}{cc}
\alpha_k(\eta_0,\eta) &\beta_k(\eta_0,\eta)\\
\beta_k^*(\eta_0,\eta) & \alpha_k^*(\eta_0,\eta)\end{array}\right)
\left(\begin{array}{c} a_{k\ell m}(\eta_0) \\  a^*_{k\ell m}(\eta_0)\end{array}\right) ,
\end{equation}
where the evolution functions $\alpha_k$ and $\beta_k$ can be obtained from the solutions (\ref{soluc}).

Standard results now tell us that  the dynamics is unitarily implementable (in the above described Fock quantization) if and only if the coefficients {$\beta_k$} are square summable, i.e. if and only if
\begin{equation}
\label{cond}
{\sum_{k=0}^\infty\sum_{\ell=0}^k\sum_{m=-\ell}^{\ell}|\beta_k(\eta_0,\eta)|^2=
\sum_{k=0}^\infty(k+1)^2|\beta_k(\eta_0,\eta)|^2<\infty},
\end{equation}
where the {degeneracy factor $(k + 1)^2$} counts the number of degrees of freedom with the same dynamics.

The fulfillment of this  condition depends on the asymptotic behavior of the coefficients  $\beta_k(\eta_0,\eta)$ for large values of $k$, which  in turn depends on the asymptotics of the Legendre functions $P^\mu_\nu$ and $Q^\mu_\nu$ for large values of  $\nu=k+1/2$. A detailed analysis carried out in Ref.~\refcite{cbmv} shows that the asymptotic behavior when $k\to \infty$ is:
\begin{equation}
\label{abeta}
\beta_k(\eta_0,\eta)= {O\left(k^{-2}\right)},\qquad \forall \eta,\eta_0.
\end{equation}
It follows that the summability condition (\ref{cond}) is satisfied for all values of $\eta_0$ and $\eta$, and therefore the dynamics is unitarily implemented in the considered Fock representation. This is in complete agreement with our general results, and explicitly disproves the claim that one cannot attain (by means of a Fock quantization) quantum unitarity of the evolution for the massless field in de Sitter space.

Crucial in this result is the scaling (\ref{rescale}) and the choice of a conjugate momentum adapted to conformal time, so that (\ref{momentum}) is satisfied. In fact, this choice of canonical pair is the only one which allows for a unitary dynamics\cite{cmov}.

\section*{Acknowledgements}
This work was supported by the Projects No. FIS2011-30145-C03-02 from Spain, CERN/FP/116373/2010 from Portugal, and DGAPA-UNAM IN117012-3 from Mexico.

\end{document}